\documentclass[aps,showpacs,prl,twocolumn]{revtex4}

\usepackage{amsmath}
\usepackage{amssymb}
\usepackage{graphicx}
\usepackage{bm}
\begin{document}

\title{Nonlocal response in plasmonic waveguiding with extreme light confinement}

\author{G.~Toscano,$^1$ S.~Raza,$^{1,2}$ W.~Yan,$^{1,3}$ C.~Jeppesen,$^1$ S.~Xiao,$^{1,3}$ M.~Wubs,$^{1,3}$ A.-P.~Jauho,$^{4,3}$ S.~I.~Bozhevolnyi,$^5$ and N.~A.~Mortensen$^{1,3}$}

\email{asger@mailaps.org}

\address{$^1$Department of Photonics Engineering, Technical University of Denmark, DK-2800 Kgs. Lyngby, Denmark\\
$^2$Center for Electron Nanoscopy, Technical University of Denmark, DK-2800 Kgs. Lyngby, Denmark\\
$^3$Center for Nanostructured Graphene (CNG), Technical University of Denmark, DK-2800 Kgs. Lyngby, Denmark\\
$^4$Department of Micro and Nanotechnology, Technical University of Denmark, DK-2800 Kgs. Lyngby, Denmark\\
$^5$Institute of Technology and Innovation, University of Southern
Denmark, DK-5230 Odense, Denmark}

\date{\today}

\begin{abstract}
We present a novel wave equation for linearized plasmonic response, obtained by combining the coupled real-space differential equations for the electric field and current density. Nonlocal dynamics are fully accounted for, and the formulation is very well suited for numerical implementation, allowing us to study waveguides with subnanometer cross-sections exhibiting extreme light confinement. We show that groove and wedge waveguides have a fundamental lower limit in their mode confinement, only captured by the nonlocal theory. The limitation translates into an upper limit for the corresponding Purcell factors, and thus has important implications for quantum plasmonics.
\end{abstract}


\maketitle Wave propagation along dielectric waveguide structures has over the years been extended
also to plasmonic systems with waveguide modes in the form of
surface-plasmon polaritons. Plasmonic waveguides have attracted
considerable attention during the past decade, primarily due to
their ability to support extremely confined modes, i.e., modes that
do not exhibit a diffraction-limited cutoff for progressively
smaller waveguide cross sections but transform themselves into their
electrostatic counterparts~\cite{Gramotnev:2010}. Investigations of
nanowire~\cite{Ditlbacher:2005}, groove~\cite{Bozhevolnyi:2005} and
wedge~\cite{Moreno:2008} waveguides, shown to ensure extreme light
confinement, raise a natural interest in the influence of nonlocal
effects on strongly confined plasmonic
modes~\cite{Garcia-de-Abajo:2008}. Waveguiding by metal
nanowires~\cite{Aers:1980} and more recently plasmonic focusing by
conical tips~\cite{Ruppin:2005,Wiener:2012} have been studied in the
context of nonlocal response. However, with the exception of few
analytical studies of simple planar
geometries~\cite{Boardman:1976,Yan:2012}, nonlocal effects in the
dispersion properties of complex waveguides remain unexplored, a
circumstance that can partly be explained by the added complexity
due to nonlocal effects as compared to the widespread framework of the
local-response approximation (LRA)~\cite{Maier:2007}.

There is also another good reason to look for nonlocal effects in
extreme light confinement. Subwavelength mode confinement implies
large effective Purcell factors and thereby strong coupling of
single emitters to nearby plasmonic waveguide modes~\cite{Chang:2006}. The latter opens a doorway to quantum
optics with surface plasmons, including the possibilities for
realization of single-photon transistors~\cite{Chang:2007} and
long-distance entanglement of qubits~\cite{Gonzalez-Tudela:2011}.
Since one would expect that the plasmonic mode confinement is
fundamentally limited by nonlocal effects, similarly to nonlocal
limits in the field enhancement of localized plasmon
excitations~\cite{Toscano:2012b,Ciraci:2012}, studies of the
plasmonic mode confinement beyond the LRA are of great interest for
quantum plasmonics. More specifically, in the LRA higher single-photon efficiencies~\cite{Chang:2006} and Purcell factors~\cite{Chang:2007} have been found to occur for smaller waveguide radii $R$, and the $R \to 0$ limit is commonly taken to estimate the strongest light-matter interactions.  Nonlocal response effects become increasingly important in this $R \to 0$ limit, which is an important motivation for our present study of nonlocal effects for highly confined plasmonic waveguides.

\begin{figure}[b!]
\centering \includegraphics[width=1\columnwidth]{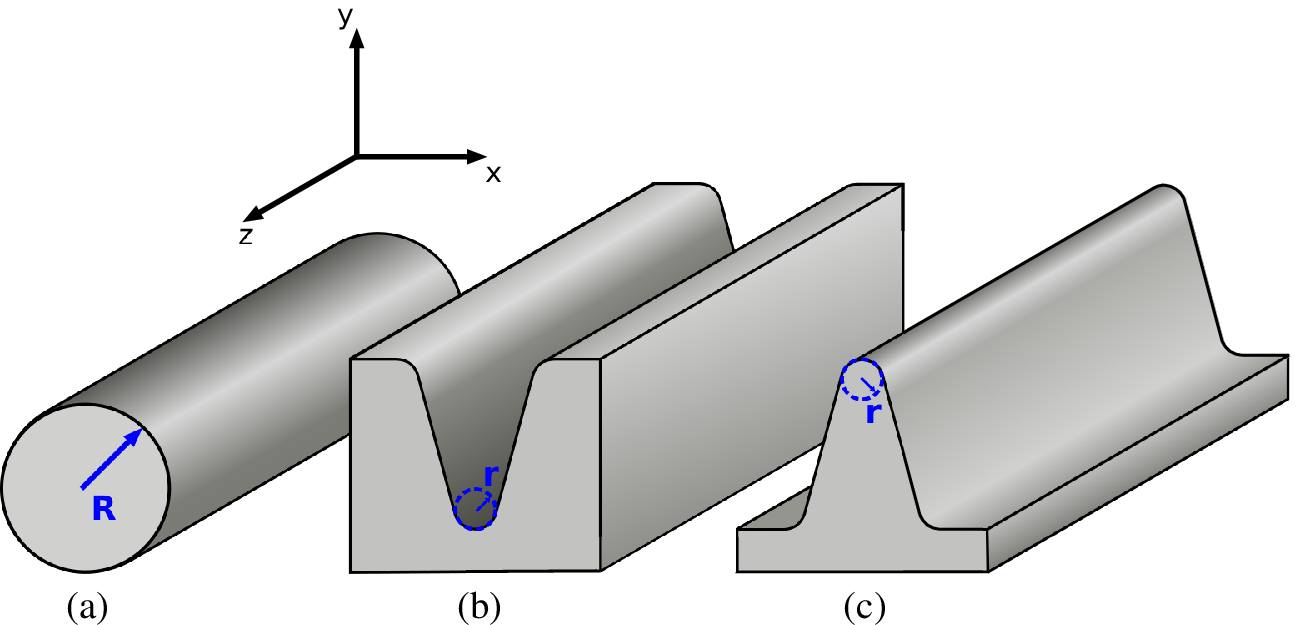}
\caption{(Color online) Generic plasmonic waveguiding geometries with wave
propagation in the $z$-direction and extreme transverse confinement
in the $xy$-plane due to subnanometer geometric dimensions, e.g.
the nanowire radius $R$ or the edge radius-of-curvature $r$.} \label{fig1}
\end{figure}

In this Letter, we derive a novel wave equation which fully takes into account the nonlocal dynamics of an often-employed hydrodynamical model (HDM). We apply the wave equation to plasmonic waveguides (Fig.~\ref{fig1}) with extreme light confinement, defined by the subnanometer
dimensions of the waveguide cross section. After stringent bench-marking of our
approach against the analytically tractable case of nanowires with circular cross-section, we analyze in detail groove and wedge waveguides and demonstrate the existence of fundamental limits in their mode confinement and Purcell factors, imposed by the nonlocal effects. At the same time, our results reveal that there is room for downsizing present-day quantum plasmonic devices before these fundamental limitations set in.

The nonlocal response, or spatial dispersion, is a consequence
of the quantum many-body properties of the electron gas, which we here
take into account within a semi-classical
model~\cite{Bloch:1933a,Barton:1979a,Boardman:1982a,Pitarke:2007a}.
In this model the equation-of-motion for an electron in an electrical field
is supplemented with a hydrodynamic pressure term
originating from the quantum kinetics of the electron gas. By
linearization, the plasmonic response is governed by the following
pair of coupled real-space differential equations~\cite{Raza:2011a}:
\begin{subequations}
\label{eq:coupledequations}
\begin{equation}
{\mathbf\nabla}\times{\mathbf\nabla}\times{\mathbf E}({\bf r})=\left(\tfrac{\omega}{c}\right)^2{\mathbf E}({\bf r}) +
i\omega\mu_0 {\mathbf J}({\bf r}),\label{eq:Maxwell}
\end{equation}
\begin{equation}
    \tfrac{\beta^2}{\omega\left(\omega+i/\tau\right)} {\mathbf \nabla} \left[ {\mathbf \nabla} \cdot {\mathbf J}({\bf r}) \right] +
 {\mathbf J}({\bf r}) = \sigma({\bf r}) {\mathbf E}({\bf r}).
\label{eq:lmotion}
\end{equation}
\end{subequations}
Here, the term ${\mathbf \nabla} \left[ {\mathbf \nabla} \cdot  {\mathbf
J}\right]={\mathbf\nabla}\times{\mathbf\nabla}\times{\mathbf J}+
{\mathbf \nabla}^2 {\mathbf J}$ is a correction to Ohm's law and
scales as $\beta^2=(3/5) v_F^2$ within the Thomas--Fermi
model~\cite{Halevi:1995} with $v_F$ being the Fermi velocity.
For simplicity we neglect here any interband
effects present in real metals; these can be included
straightforwardly~\cite{Toscano:2012a,supplemental}. In our numerical solutions we will consider Drude parameters appropriate for silver~\cite{Rodrigo:2008}. Assuming a hard-wall confinement associated with a high
work function, the boundary conditions for the current at the metal
surface become particularly simple: the tangential component is
unrestricted while the normal component vanishes due to the current
continuity and vanishing of all electron wave functions at the
surface~\cite{Raza:2011a,Yan:2012}.

For analytical progress one can eliminate the current from Eq.~(\ref{eq:Maxwell}), thereby arriving at an integral equation where a dyadic Green's function accounts for the nonlocal dynamics of the electron gas~\cite{Mortensen:2012,Mortensen:2013}. Alternatively, the coupled equations (\ref{eq:Maxwell}) and (\ref{eq:lmotion}) form a natural starting point for a numerical treatment of arbitrarily shaped metallic nanostructures, e.g., with a state-of-the-art finite-element method~\cite{Toscano:2012a,Hiremath:2012}. Recently, we employed this approach to study field enhancement and SERS in groove structures~\cite{Toscano:2012b}. However, for waveguiding geometries
we seek solutions of the form ${\mathbf E}({\bf r})\propto \exp(ik_z z)$ leading to an eigenvalue problem for $k_z(\omega)$ with a six-component eigenvector $\{{\mathbf E},{\mathbf J}\}$. In that context the coupled-equation formulation is numerically less attractive. Here, instead, we  eliminate the current from Eq.~(\ref{eq:lmotion}), a procedure that, after straightforward manipulations using standard vector calculus~\cite{supplemental}, results in an appealingly compact, but yet entirely general nonlocal wave equation:
\begin{subequations}\label{eq:Maxwell_NL}
\begin{equation}
{\mathbf\nabla}\times{\mathbf\nabla}\times{\mathbf E}({\bf
r})=\left(\tfrac{\omega}{c}\right)^2\hat{\varepsilon}_{\rm
\scriptscriptstyle NL}({\bf r}){\mathbf E}({\bf r}),
\end{equation}
\begin{equation}\label{eps-op}
\hat{\varepsilon}_{\rm \scriptscriptstyle NL}({\bf
r})=\varepsilon_{\scriptscriptstyle D}({\bf
r})+\tfrac{\beta^2}{\omega\left(\omega+i/\tau\right)}\nabla^2.
\end{equation}
\end{subequations}
Here, the operator $\hat\varepsilon_{\rm \scriptscriptstyle NL}({\bf r})$ contains the nonlocal effects.  In the limit $\beta\to 0$, $\hat\varepsilon_{\rm \scriptscriptstyle NL}({\bf r})$ reduces to the usual Drude dielectric function $\varepsilon_{\scriptscriptstyle D}({\bf r})=1+i\sigma({\bf r})/(\varepsilon_0\omega)=1-\omega_p^2({\bf r})/[\omega(\omega+i/\tau)]$ used in the LRA. Thus, with a simple rewriting we have turned the coupled-wave equations into a form reminiscent of the usual wave equation, with all aspects of nonlocal response contained in the Laplacian term $\beta^2\nabla^2$ in $\hat{\varepsilon}_{\rm \scriptscriptstyle NL}({\bf r})$. This is the main theoretical result of this Letter. In passing, we note that with Eq.~(\ref{eps-op}) we immediately recover the dispersion relation $\omega(k)=\sqrt{\omega_p^2+\beta^2k^2}$ for bulk plasmons in translationally invariant plasma~\cite{supplemental}. Clearly, the single-line form is beneficial for the conceptual
understanding and further analytical progress, as well as for numerical implementations: the additional Laplacian does not add any complications beyond those already posted by the double-curl operator on the left-hand side equation. Likewise the  boundary condition that was imposed on the current ${\mathbf J}$ in Eq.~(\ref{eq:coupledequations}) translates into an additional boundary condition on the electric field in Eq.~(\ref{eq:Maxwell_NL}), see~\cite{supplemental}. While Eq.~(\ref{eq:coupledequations}) can be solved numerically for scattering problems~\cite{Toscano:2012a,Hiremath:2012,Toscano:2012b} and some waveguide problems~\cite{Huang:13}, the result in Eq.~(\ref{eq:Maxwell_NL}) is clearly a major advancement for efficient and accurate numerical eigenvalue solutions in waveguiding geometries with arbitrarily shaped waveguide cross sections. In particular, differential operations reduce to a Laplacian and the dimension of the eigenvalue problem is reduced from six field components to only three.

We now apply the developed formalism to the waveguide configurations of Fig.~\ref{fig1} which can provide extreme light confinement~\cite{Gramotnev:2010}: \emph{i}) metal nanowires with circular cross sections~\cite{Ditlbacher:2005} where analytical solutions~\cite{Ruppin:2005} are available for benchmarking of the numerics, \emph{ii}) grooves in metal~\cite{Bozhevolnyi:2005}, and \emph{iii}) metal wedges~\cite{Moreno:2008}. In addition to the usual mode characteristics, effective index and propagation length, we also evaluate the effective mode area: $A_{\scriptscriptstyle\rm eff}=V_{\scriptscriptstyle\rm eff}/L$, where $V_{\scriptscriptstyle\rm eff}$ is the effective mode volume associated with the Purcell effect, i.e.,
\begin{equation}
\label{eq3} A_{\scriptscriptstyle\rm eff}=\frac{\int_{V_{m}+V_{a}}
\mbox{d}x\mbox{d}y\, u({\bf r})}{\max_{V_{a}}\big\{ u({\bf r})
\big\}},
\end{equation}
where $u({\bf r})$ is the electromagnetic energy
functional~\cite{supplemental}. The cross-sectional integral extends
over the volumes $V_m$ and $V_a$ occupied by metal and air,
respectively, while the evaluation of the maximal energy density is
restricted to the air region where dipole emitters can be placed.

\begin{figure}[t!]
\centering \includegraphics[width=1\columnwidth]{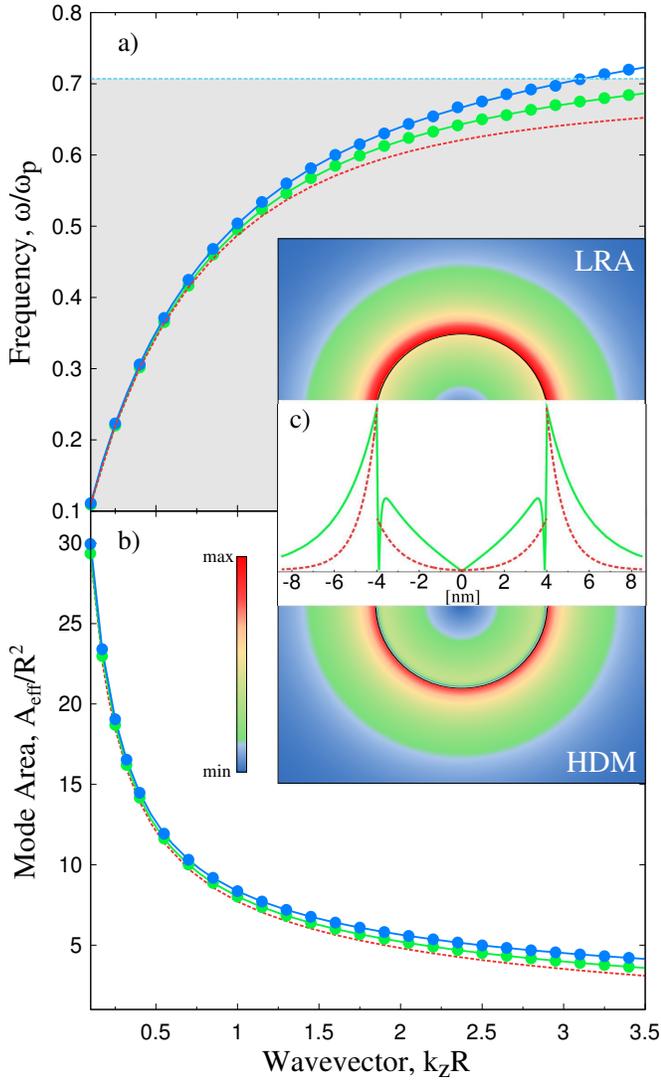}
\caption{(Color online) Fundamental waveguide mode of a cylindrical silver nanowire
embedded in air. (a) Dispersion relation $\omega(k_z)$ and (b)
normalized effective mode area within the HDM for the nanowire
radius $R=2$\,nm (blue) and $4$\,nm (green), respectively, showing
excellent agreement between numerical solutions of
Eq.~(\ref{eq:Maxwell_NL}) (solid points) and analytical results
(solid lines). For comparison, the red-dashed curve shows the
universal result of the nonretarded LRA, with its large-$k_z$
limiting value of $\omega_p/\sqrt{2}$ indicated in (a) by the
horizontal line. (c) Radial distribution of the electric field
$\big|{\mathbf E}_{\rho}\big|$ at $\omega=0.6 \, \omega_{\rm p}$ for
$R=4$\,nm, contrasting the continuous field variation in the HDM
with its usual boundary discontinuity in the LRA.} \label{fig2}
\end{figure}

The dispersion curves and effective mode areas (normalized to the
nanowire cross section) calculated for silver nanowires of different
radii [Fig.~\ref{fig2}(a,b)] exhibit a blueshift and increased mode area (for fixed $k_z$) when
taking nonlocal effects into account. The numerical results of Eq.~(\ref{eq:Maxwell_NL}) show excellent agreement
with the corresponding analytical results previously derived from Eq.~(\ref{eq:coupledequations})~\cite{Ruppin:2005}.
Importantly, nonlocal dynamics influences strongly the mode field
distribution [see Fig.~\ref{fig2}(c)], because, contrary to the LRA
case, the normal component of the electrical field within the HDM is
continuous across the interfaces (this is a special case for a Drude metal without interband effects and surrounded by vacuum~\cite{supplemental}). It is indeed
seen [Fig.~\ref{fig2}(c)] that $|{\mathbf E}_{\rho}|$ is
discontinuous on the boundary in the local case, while it varies
continuously across the boundary in the nonlocal case. This
variation occurs in a region extending over $\approx 0.1$\,nm, that
is of the order of the Fermi wavelength of silver.

The results for cylindrical nanowires, while demonstrating the main effects of nonlocal dynamics on the mode characteristics, indicate that the quantitative changes are modest even for very small radii (Fig.~\ref{fig2}). In order to explore \emph{fundamental} limitations, one has to consider the limit of vanishing radii of curvature. While subnanometer radii appear unrealistic for nanowires, fabrication of grooves cut in metal and metal wedges, e.g., by nanoimprint lithography~\cite{Nielsen:2008}, can in fact result in nm-sharp edges with corresponding nm-sized wedge modes~\cite{Moreno:2008}. We expect that nonlocal effects then come into play.

\begin{figure}[b!]
\centering \includegraphics[width=1\columnwidth]{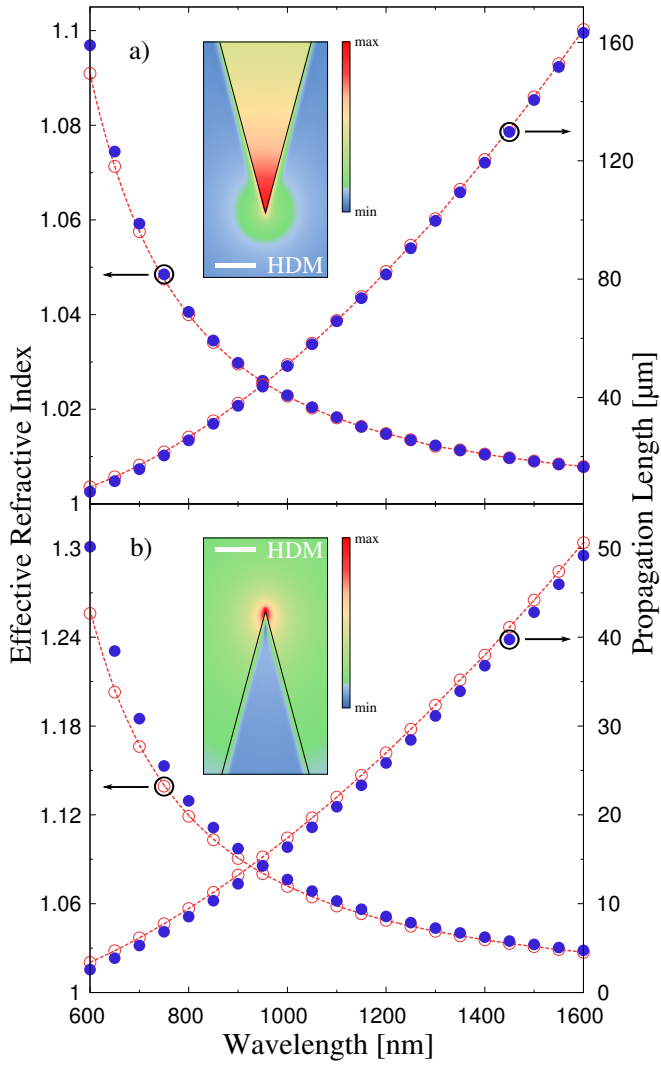}
\caption{(Color online) Effective index (left axis) and propagation length (right
axis) versus wavelength for the fundamental mode in complimentary
(a) V-groove and (b) wedge silver waveguides, both with an opening
angle of $30^{\circ}$. The nonlocal results (solid circular
symbols) obtained with Eq.~(\ref{eq:Maxwell_NL}) are contrasted to
the LRA (open circles), with dashed lines serving as eye guides.
Results for mathematically sharp structures with $r=0$ (blue solid circles)
are contrasted to finite rounding with $r=1$\,nm (red open circles).
Insets show field-intensity distributions (white scale bars are
1\,nm long) calculated within the HDM ($\lambda=600$\,nm) for
infinitely sharp edges. The fingerprint of nonlocal effects is clearly visible as the field
penetrates into the metal by a distance of the order of the Fermi
wavelength of silver.} \label{fig3}
\end{figure}

Rather surprisingly, the mode effective index and propagation length
calculated for silver grooves and wedges (Fig.~\ref{fig3}) exhibit
even weaker influence of the nonlocal effects as compared to the case
of nanowires (Fig.~\ref{fig2}). In fact, there is no noticeable
difference between the LRA- and HDM-based results obtained for
1-nm-radius of edges. In the limit of mathematically sharp edges, the
mode effective index becomes only slightly larger and the
propagation length slightly smaller than those calculated for 1\,nm
edge radius (Fig.~\ref{fig3}). We explain this result by the fact
that groove and wedge plasmonic modes are only partially affected by
the very tip, being distributed also and predominantly over flat
edges (see insets in Fig.~\ref{fig3}).

\begin{figure}[t!]
\centering \includegraphics[width=1\columnwidth]{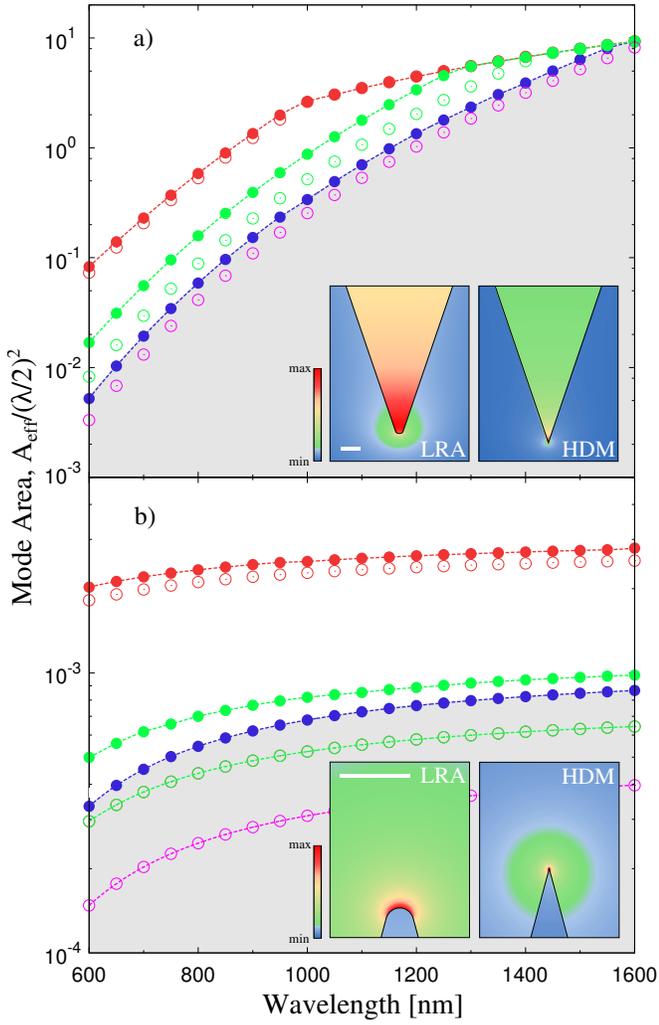}
\caption{(Color online) Normalized mode area versus wavelength for the fundamental
mode in complimentary (a) V-groove and (b) wedge silver waveguides,
both with opening angles of $30^{\circ}$. The HDM results (solid symbols) are contrasted to
the LRA (open circles) for $r=1$\,nm (red) and $r=0.2$\,nm (green). Results for mathematically sharp structures with $r=0$ (blue solid circles) define a lower limit in the HDM (grey-shaded regions are inaccessible). For the LRA, the $r=0.1$\,nm results (magenta) exceed this limit and the mode area tends to zero when $r
\rightarrow 0$. Insets show field-intensity distributions (white scale bars are 5\,nm long)
at $\lambda=600$\,nm. The LRA intensities are with rounding
$r=1$\,nm, while $r=0$ is used for the HDM maps.} \label{fig4}
\end{figure}

The situation changes drastically when one considers the mode confinement, using the mode area associated with the Purcell factor, Eq.~(\ref{eq3}). We recall that the field enhancement calculated within the LRA grows without bound for progressively sharper pointed structures while it remains finite when calculated within
HDM~\cite{Toscano:2012b,Ciraci:2012}.  Analogously, in the present case, one may expect that the mode area calculated within the LRA decreases without bound for a decreasing edge radius, while it may saturate within the HDM. LRA-based simulations for subnanometer radii of edges show (Fig.~\ref{fig4}) that the mode area indeed tends to zero, without any apparent saturation. This trend is more pronounced for wedges than for grooves, because the wedge geometry ensures generally a better mode confinement [cf. Figs.~\ref{fig4}(a) and (b)], as was also noted previously~\cite{Moreno:2008}. At the same time, the simulations conducted within the HDM demonstrate clearly the existence of a lower bound for the mode area which remains finite even for mathematically sharp edges (blue circles in  Fig.~\ref{fig4}). The associated Purcell factors can be estimated by inverse of the normalized mode areas displayed in Fig.~\ref{fig4}~\cite{Oulton:2008}. Thus, our calculations show that there is a fundamental limit for the maximum Purcell factors achievable with plasmonic waveguides. It is interesting that the upper limit of Purcell factors evaluated in this way decreases noticeably in the long-wavelength regime. This feature is related to a general weakening of all plasmonic effects, including waveguiding~\cite{Gramotnev:2010}, for longer wavelengths (with metals approaching the limiting case of perfect conductors). At the same time, in the case of wedges, these factors remain substantial
even at telecom wavelengths, with the propagation lengths becoming considerably long (Fig.~\ref{fig3}) and amenable for circuitry application. It should also be borne in mind that the plasmonic field confinement in both grooves and wedges increases for smaller opening angles ~\cite{Bozhevolnyi:2005,Moreno:2008}, so that even larger Purcell factors can be achieved, albeit at the expense of shorter propagation.

In conclusion, using a novel wave equation accounting for nonlocal dynamics, we considered plasmonic waveguides with extreme light confinement and demonstrated the existence of a fundamental limit in their mode confinement imposed by nonlocal effects. Our results imply fundamental limitations in the corresponding Purcell factors, showing at the same time the possibility of achieving very high Purcell factors with V-groove and wedge waveguides that ensure sufficiently long propagation lengths for applications in quantum plasmonics. Here, we have focused on single-connected metal geometries where dominating currents are naturally of an Ohmic nature, whereas tunneling currents may cause important limitations too in e.g. closely spaced metallic objects~\cite{Esteban:2012}. Finally, beyond the linear response fundamental limitations may arise due to nonlinearities~\cite{Ginzburg:2010}.

\emph{Acknowledgments.} This work was financially supported by an H.~C. {\O}rsted Fellowship (W.Y.) and the Center for Nanostructured Graphene (CNG) is sponsored by the Danish National Research Foundation, Project DNRF58.

\newpage

\end{document}